\documentclass[twocolumn,preprintnumbers,amsmath,amssymb,floats]{revtex4}
\usepackage{graphicx}
\usepackage{dcolumn}
\usepackage{bm}
\newcommand{\be}{\begin{equation}}
\newcommand{\ee}{\end{equation}}
\newcommand{\bea}{\begin{eqnarray}}
\newcommand{\eea}{\end{eqnarray}}
\newcommand{\no}{\nonumber \\}
\newcommand{\etal}{{\it et al.}~}
\newcommand{\eg}{{\it e.g.}}

\def\la{\langle}
\def\ra{\rangle}
\def\H#1{{}^{#1}\mbox{H}}
\def\He#1{{}^{#1}\mbox{He}}
\def\MeV{{\mbox{MeV}}}
\def\gA{g_A}
\def\fpi{f_\pi}

\def\ttz{(\tau_1\times\tau_2)^z}

\def\fpi{f_\pi}

\def\TtimesT{{{\hat T}_T^{(\times)}}}

\def\hatr{{\hat r}}

\def\vJ{{\bm J}}

\def\vj{{\bm j}}
\def\vr{{\bm r}}
\def\vq{{\bm q}}

\def\vs{{\bm \sigma}}

\def\vtau{{\bm \tau}}

\def\nlo#1{\mbox{N$^{#1}$LO}}
\def\oforderof#1{{\cal O}\left({#1}\right)}
\begin{document}

\preprint{KIAS-P03082}
\title{ $Hen$ Process in Effective Field Theory}
\author{Young-Ho Song$^{(a)}$ and Tae-Sun Park$^{(b)}$}
\affiliation{$(a)$ Department of Physics, Seoul National University, Seoul 151-742, Korea}
\affiliation{$(b)$ School of Physics, Korea Institute for Advanced Study, Seoul 130-012, Korea}

\date{\today}
\begin{abstract}
An effective field theory technique that combines the standard
nuclear physics approach and chiral perturbation theory is applied
to the $hen$ process, ${}^{3}\mbox{He}+n\rightarrow {}^{4}\mbox{He}+ \gamma$.
For the initial and final nuclear states,
high-precision wave functions are generated via the variational
Monte Carlo method using the Argonne $v_{14}$ potential and Urbana VIII  trinucleon
interactions, while the relevant transition operators are
calculated up to ${\cal O}(Q^4)$ in HB$\chi$PT.
The imposition of the renormalization condition
that the magnetic moments of ${}^{3}\mbox{He}$ and ${}^{3}\mbox{H}$ be reproduced
allows us to carry out a parameter-free calculation of the $hen$ cross section.
The result, $\sigma=(60\pm 3\pm 1)~\mu b$, is in reasonable agreement with the
experimental values,  $(54\pm 6)~\mu b$
and $(55\pm 3)~\mu b$. This agreement demonstrates
the validity of the calculational method previously used
for estimating the reaction rate of the solar $hep$ process.

\end{abstract}

\maketitle

One of the important advantages of using effective field theory (EFT) in nuclear
physics is that it leads to well-controlled predictions
for nuclear observables over which
the conventional model approaches have little
or no control.
One such case is the
$hep$ process $\He3+p\rightarrow \He4+\nu_e+e^+$,
which can be potentially important for solar neutrino physics
and non-standard physics in the neutron sector.
In a recent paper, Park {\it et al.}~\cite{hep}
have developed an EFT approach,
to be referred to as ``MEEFT"\cite{BR03},
and succeeded in making a
totally parameter-free calculation of the $hep$ cross section.
In MEEFT, the relevant transition operators are
systematically derived using chiral perturbation theory,
while the initial and final nuclear wave functions are
generated using the standard nuclear physics approach (SNPA).
Here SNPA refers to an approach in which the wave functions of light
nuclei (mass number A) are obtained by solving exactly
the A-body Schr\"{o}dinger equation
with a Hamiltonian that contains
a high-precision phenomenological potential.
With the use of MEEFT,  Park {\it et al.} arrived at a prediction
of the $hep$ $S$-factor with $\sim 15\%$ accuracy.
Given the fact that the predicted values of the $hep$ $S$-factor
ranged over two orders of magnitude in the past,
the result obtained in \cite{hep} could be
considered a remarkable feat.
It is to be noted, however, that the $\sim 15\%$ uncertainty
associated with the $hep$ $S$-factor
(although acceptable on its own merit) is strikingly larger than
the uncertainty associated with the $S$-factor for the $pp$ fusion process,
the latter estimated to be less than 0.5\% \cite{hep}.
%
There are two plausible causes for such a
``large" uncertainty in the $hep$ calculation.
First, since the calculation in \cite{hep}
relies entirely on the information provided by the two-body
and three-body observables, it may miss part of
the dynamics which is intrinsically four-body in nature.
Secondly, symmetry mismatch between the initial and final states of the
$hep$ process, a feature that is absent in the two- and three-body systems,
induces a strong suppression of the otherwise dominant single-particle
contribution, resulting in an enhancement of higher-order corrections,
perhaps including four-body interactions.

The purpose of this Letter is to test the reliability of MEEFT
used to compute the $hep$ process by applying the same
method to the $hen$ process
\be
\He3 +n \rightarrow \He4 + \gamma
\label{henprocess}
\ee
at threshold;
$hen$ is a four-body process
that has close similarity to $hep$ in that
the pseudo-orthogonality of the initial and final wave functions
strongly suppresses the leading-order one-body contribution.
In fact, the degree of suppression for $hen$ is such that
the exchange current ``corrections" become dominant terms.
Meanwhile, since the experimental value of the threshold $hen$ process
is known with reasonable accuracy:
$\sigma_{exp}=(54\pm 6)~\mu b$ \cite{Wolfs} and
$\sigma_{exp}=(55\pm 3)~\mu b$ \cite{wervelman},
$hen$ provides a good testing ground for the validity of MEEFT.
It is to be noted that these experimental values have never been
explained satisfactorily before.
We remark, in particular, that
very elaborate SNPA calculations
by Carlson \etal\cite{hen1} and by Schiavilla \etal\cite{hen2}
give $\sigma \!=\!112~\mu b$ and
$\sigma\!=\!86~\mu b$, respectively.
It has been noted that $\sigma$
is extremely sensitive to the $\He3+n$ scattering length,
$a_n$; the use of the updated value,
$a_n^{\rm exp}=3.278(53)$ fm \cite{zimmer}, makes
the calculated of $\sigma$ even larger.
In view of the high quality of the wave functions used
in these calculation,
it is very probable that the problem lies in the many-body currents used.
This observation makes the use of MEEFT
for $hen$ all the more interesting.

Our parameter-free calculation based on MEEFT gives
for the threshold $hen$ cross section
\be \sigma=(60.1\pm 3.2\pm 1.0)~\mu b, \label{result}\ee
which agrees with the above-quoted experimental values.

The strategy of MEEFT, explained in
\cite{hep,park2001}, is close in spirit to Weinberg's original
scheme~\cite{weinberg} based on the chiral expansion of
``irreducible terms". MEEFT has been used with great success to
describe the highly suppressed isoscalar amplitude pertaining to
the $n+p\rightarrow d+\gamma$ process~\cite{excV,isoS}
and many weak processes (for a recent review, see
Ref. \cite{kubodera}).
In MEEFT, the relevant current operators are derived
systematically by applying heavy-baryon chiral
perturbation theory (HB$\chi$PT)
up to a specified order.
The nuclear matrix elements are then evaluated by
sandwiching these current operators
between the high-precision wave functions
obtained in SNPA.
In EFT, short-distance physics is represented as contact
counter-terms, and the coefficients of these terms,
called the low-energy constants (LECs)
are determined by requiring that a selected set
of experimental data be reproduced.
This renormalization procedure is expected to remove
to a large extent the model-dependence of the short-range behavior
of the adopted phenomenological wave functions,
and this feature is very important for making
model-independent predictions.
We should note here that the strategy employed here is closely
related to that used for the universal $V_{low-k}$ in the recent
renormalization-group approach to nuclear interactions~\cite{stonybrook}.

The total cross section of the $hen$ reaction reads
\be
\sigma=\frac{2\alpha\omega}{v_n}\frac{1}{4}\sum_{M=-1}^{1}
|\la\He4|{\bf j}_{T}(\vq)|n+\He3;1M\ra|^2, \ee
where $\alpha$ is
the fine structure constant, $(\omega,\, \vq)$ is the
four-momentum of the outgoing photon with $|\vq|=\omega\simeq
20.521\ \MeV$, $v_n$ is the velocity of the incident thermal
neutron, $M$ is the spin quantum number of the initial system, and
$\vj_T$ is the electromagnetic (EM) current operator transverse to
the photon direction. As the process is isotropic at threshold,
we can choose ${\hat q}={\hat z}$.

The formalism employed here
is essentially the same as in the $hep$ process, so we refer
to Ref.\cite{hep} whenever appropriate.
We obtain the EM current using HB$\chi$PT,
which contains the nucleons and pions as explicit degrees of freedom,
with all the other massive fields integrated out.
In HB$\chi$PT, operators
carry
a typical momentum scale of the process
and/or the pion mass,
which is regarded as small compared to the
chiral scale
$\Lambda_\chi\sim 4\pi f_\pi \sim m_N \sim 1$~GeV,
where $f_\pi\simeq 93$ MeV is the pion decay constant and
$m_N\simeq 939$ MeV the nucleon mass.
The leading
space-components of the EM currents are of order of $Q$, and hence we
refer to terms of order $Q^{n+1}$ as $\nlo{n}$. In this
paper, we shall limit ourselves up to $\nlo{3}$ or
$\oforderof{Q^4}$.
It is important to take
into account the kinematical suppression of the time component of
the nucleon four-momentum and the smallness of the photon
energy/momentum. For simplicity, we count the time component
of the nucleon four-momentum $q_0$ and $|\vq|$ as $Q^2$ order
rather than $Q$. {\it It is important
to emphasize that, up to $\nlo3$,  three-body and four-body
current operators do not appear.} This means that
the nuclear systems with $A=2, 3$ and $4$ can all be
described by the $same$ current operators up to this order.
%
%
%
%

The leading order one-body (1B) currents are given by
\be
\vJ_i=e^{-i\vq\cdot\vr_i}\left[\frac{(1+\tau^z_i)}{2 m_N}{\bar p_i}
            +\frac{(\mu_S+\mu_V\tau^z_i)}{4m_N}i\vq\times\vs_i\right]
\ee
 where, ${\bar p}\equiv
\frac{1}{2}(i\stackrel{\leftarrow}{\nabla}-i\stackrel{\rightarrow}{\nabla})$
acts only on the wave function.
Relativistic corrections to the 1B currents are also included
in the calculation.
It should be noted
that the two-body (2B) currents in momentum space are valid only up to
a certain cutoff $\Lambda$. This implies that the currents should be appropriately regulated.
This is the key point in our approach. This cutoff defines the
energy/momentum scale of EFT below which reside the chosen
explicit degrees of freedom. We use a Gaussian regulator in
performing the Fourier transformation from momentum to
coordinate space. All the 2B currents can be written in the form
of $\vJ_{12}(\vr_1,\vr_2)=e^{-i\vq\cdot (\vr_1+\vr_2)/2}\vj_{12}(\vr_1-\vr_2)$.
The $\nlo{1}$ correction is due to the {\em soft}-one-pion-exchange,
\bea
&\vj_{12}^{1\pi}(\vr)=-\frac{g_A^2 m_\pi^2}{12 f_\pi^2}\vtau_{\times}^z\vr
          [\vs_1\cdot\vs_2 {\bar y}_{0\Lambda}^\pi(r)+S_{12}y_{2\Lambda}^\pi(r)]
          \no
       &+i\frac{g_A^2}{8f_\pi^2}\vq\times
         [{\hat T}_{S}^{(\times)}(\frac{2}{3}y_{1\Lambda}^\pi(r)-y_{0\Lambda}^\pi(r))
         -{\hat T}_{T}^{(\times)}y_{1\Lambda}^\pi(r)],
\label{1pi}\eea
where $g_A=1.2601$, $S_{12}= 3 \vs_1\cdot
\hatr\,\vs_2\cdot \hatr - \vs_1\cdot\vs_2$;
we have also defined
${\hat T}_{S,12}^{(\odot)}\equiv \vtau_\odot^z \vs_\odot$
and  ${\hat T}_{T,12}^{(\odot)}\equiv
\vtau_\odot^z\left[{\hat r}\,{\hat
r}\cdot\vs_\odot-\frac13\vs_\odot\right]$,
$\vtau_{\odot}=\vtau_1\odot\vtau_2$,
$\vs_{\odot}=\vs_1\odot\vs_2$,
$\odot= \pm,\, \times$.
The explicit forms of
the regulated Yukawa functions are given in Ref.\cite{hep}.

There are no corrections at $\nlo{2}$.
At $\nlo{3}$,
there occur vertex corrections to the
one-pion exchange $j^{1\pi C}$,
the two-pion-exchanges $j^{2\pi}$, and the contact counter-term contributions $j^{\rm CT}$.
The $j^{1\pi C}$ investigated
in detail in Ref.~\cite{excV,isoS} reads
\bea
&\vj_{12}^{\rm 1\pi C}= i \vq \times \big\{
-\frac{\gA^2}{8 \fpi^2}
  ({\bar c}_\omega+{\bar c}_\Delta)
  \big[({\hat T}_{S}^{(+)}+ {\hat T}_{S}^{(-)}) \frac{{\bar y}^\pi_{0\Lambda}}{3}
\no
  &+({\hat T}_{T}^{(+)} + {\hat T}_{T}^{(-)})\,y^\pi_{2\Lambda}\big]
-\frac{\gA^2 }{16 \fpi^2} {\bar c}_{\Delta}\TtimesT y^\pi_{2\Lambda}
\no
&+\frac{1}{16 f_\pi^2}{\bar N}_{WZ}\vtau_1\cdot\vtau_2
  \big[\vs_{+}{\bar y}^\pi_{0\Lambda}
+(3\hatr\hatr\cdot\vs_{+}-\vs_{+}) y^\pi_{2\Lambda}\big]\big\},
\label{1piC}\eea
where
$\bar c_\omega\simeq 0.1021$, $\bar c_\Delta\simeq 0.1667$, and
${\bar N}_{WZ}\simeq 0.02395$.
Although $j^{1\pi C}$ contains the so-called ``fixed terms"
needed to ensure Lorentz covariance,
we neglect them here since
they are expected to be small.
The two-pion-exchange diagrams,
except for the box-diagrams, can be found in ref.\cite{excV}.
Adding the box-diagram contributions and subtracting the reducible
iterated one-pion-exchange part, we obtain
\bea
&&\vj_{12}^{2\pi}=-\frac{1}{128\pi^2 f_\pi^4} \Big\{
\ttz\hatr \frac{d}{dr} L_0
\no
&&- i \vq \times\left[({\hat T}_{S}^{(+)}- {\hat T}_{S}^{(-)} ) L_S
+ ({\hat T}_{T}^{(+)}- {\hat T}_{T}^{(-)}) L_T\right] \Big\}
\label{2pi}\eea
with
\bea
L_0 &=& 2 K_2 + g_A^2 (16K_2-2K_1-2K_0)
\no && \ \ \
        -g_A^4 (16 K_2+5 K_1+5 K_0) + g_A^4 \frac{d}{dr} (r K_1),
\no
L_S &=& \frac{g_A^2}{3} r \frac{d}{dr} K_0  + \frac{g_A^4}{3} \big[
4 K_1 -2 K_0+r\frac{d}{d r} (K_0 + 2 K_1) \big],
\no
L_T&=&-\frac{g_A^2}{2}  r\frac{d}{d r} K_0+\frac{g_A^4}{3}
\big[ 4 K_{\rm T} - r \frac{d}{d r} (K_0 + 2 K_1) \big].
\eea
%
The loop functions $K's$ are defined in
Ref. \cite{hep,excV}. Finally there are contact
contributions of the form
 \be \vj_{12}^{\rm CT}=\frac{i}{2m_p}\vq\times
         [g_{4S}(\vs_1+\vs_2)+g_{4V}T_S^{(\times)}]\delta^{(3)}_\Lambda(\vr),
 \label{CT}\ee
where $g_{4S}=m_p g_4$ and
$g_{4V}=- m_p (G_A^R+\frac{1}{4}E_T^{V,R})$, $m_p$ is the proton mass;
$g_4$, $G_A^R$ and $E_T^{V,R}$ are the coefficients of the counter-terms
introduced in \cite{isoS,excV}.
A highly noteworthy point is that,
although there are many possible counter-terms,
only two linear combinations of them are independent
for the M1 transition.
This reduction of the effective number of counter terms is due to
the Fermi-Dirac statistics; a similar reduction
has been noted for $hep$ ~\cite{hep},
where only one linear combination of LECs needs to be retained.

As the cutoff $\Lambda$ has a
physical meaning, its choice is not arbitrary.
Obviously, $\Lambda$ greater than
$\Lambda_\chi$ should be excluded.
Furthermore, $\Lambda$ should not be larger
than the mass of the
vector mesons that have been integrated out;
thus, $\Lambda \lesssim 800$ MeV.
Meanwhile, since the pion is an explicit degree of freedom in our
scheme, $\Lambda$ should be much larger than the pion mass in order to
ensure that all pertinent low-energy contributions are properly
included. In the present work,
we adopt $\Lambda=$ 500, 600 and 800 MeV
as representative values.

\begin{figure}
\includegraphics{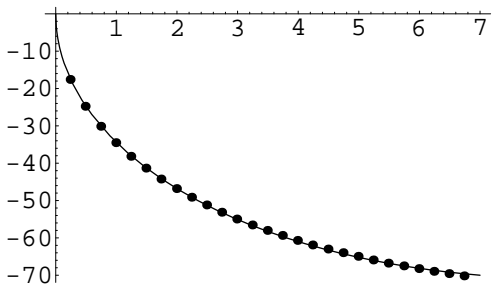}
\caption{\label{fig:phaseshift} The $\He3+n$ S-wave phase shift in degrees
as a function of
the center-of-mass energy in MeV.
The solid line gives our results obtained with the Woods-Saxon potential,
while the full dots represent the experimental data obtained from an
$R$-matrix analysis \cite{fofman}.}
\end{figure}

We use realistic variational Monte-Carlo wave functions obtained
for the Argonne v14 two-nucleon interaction and the Urbana VIII
trinucleon interactions; these are the potentials used in the previous SNPA
calculation\cite{hen1}. The asymptotic behavior of the $\He3+n$
system has been taken into account by means of the Woods-Saxon
(WS) potential, $V_{WS}(r)=V_{0}
\left(1+e^{(r-R_{0})/d_{0}}\right)^{-1}$.
We can determine the parameters in the WS
potential so as to reproduce the experimental data for the
scattering length \cite{zimmer} and the low-energy phase
shifts \cite{fofman}; the results are
$V_{0}=8.63637$ MeV, $R_{0}=4.14371$ fm and $d_{0}=0.8$ fm.
It should be pointed out \cite{hen1} that
a small difference $a_n$ can affect the $hen$ cross
section significantly.
For this reason the
experimental data are directly encoded into theory.
The resulting phase shifts are shown in figure
\ref{fig:phaseshift}.
The precise measurement of the
$\He3+n$ scattering length enables us to limit the uncertainty
associated with the scattering length to $\lesssim 1\%$ in the matrix
elements. The nuclear matrix elements of the currents
are evaluated by taking 1 million samplings in the
Metropolis algorithm.

For each value of $\Lambda$, we adjust the values of the two
independent LECs (denoted by $g_{4s}$ and $g_{4v}$)
so as to reproduce the experimental values
of the $\H3$ and $\He3$ magnetic moments.
The results are
$\{g_{4s},\,g_{4v}\}= \{1.079(9) ,\, 2.029(6)\}$,
$\{1.277(12),\,0.981(7)\}$ and
$\{1.856(22),\,-0.235(12)\}$
for $\Lambda=500$, 600 and 800 MeV, respectively.
Once the values of $g_{4s,4v}$ are fixed,
we can predict the $M1$ transition
amplitude for $hen$ without any unknown parameters.

In Table \ref{tab:matrixelement}, we list the contributions
of various terms
to the magnetic moment of $\H3$ and $\He3$ and the
$hen$ matrix elements
$\la j \ra_{hen}\equiv \frac12 \la\He4| (j_x - i j_y) |n+\He3;11\ra$,
calculated for $\Lambda=600$ MeV.
\begin{table}
\caption{\label{tab:matrixelement} The magnetic moment of
$\H3$ and $\He3$ (in units of $\mu_N$), and the $hen$
matrix elements $\la j\ra_{hen}$
(in units of $10^{-3}$ fm$^{3/2}$), calculated for $\Lambda=600$ MeV.
The numbers in the parentheses represent Monte-Carlo statistical errors.}
\begin{ruledtabular}
\begin{tabular}{l|cc|r}
       & $\mu(\H3)$ & $\mu(\He3)$ & $\la j\ra_{hen}$  \\
\hline
1B                 & $2.6068(1)$ & $-1.7839(1) $&  $-1.76(1)$   \\
\hline
2B($1\pi+ 1\pi C$) &  $0.2676(3)$ & $-0.2975(3)$ &  5.89(3)  \\
2B($2\pi$)         & $0.0334(1)$ & $-0.0334(1)$ &  0.90(1)   \\
2B(CT)             &  0.0710(3) &  $-0.0128$(3) &  0.35(1)  \\
2B(total)          &   0.3721(1) & $-0.3437(1)$ & 7.14(3)  \\
\hline
1B+2B               &   2.9789 & $-2.1276$ & 5.39(3)  \\
\end{tabular}
\end{ruledtabular}
\end{table}
The row labelled ``2B($1\pi+ 1\pi C$)" gives
contributions from eqs.(\ref{1pi},\ref{1piC}),
``2B($2\pi$)" from eq.(\ref{2pi}),
and ``2B(CT)" from eq.(\ref{CT});
``2B(total)" represents the sum of all these contributions.
As mentioned, the 1B contribution is $strongly$ suppressed
by the pseudo-orthogonality of the wave functions,
and as a result the matrix element is dominated by the
$\la \mbox{2B}\ra$.
Furthermore, since the 2B and 1B contributions have opposite signs,
there is destructive interference between them,
which, as in the $hep$ case,
makes it a more challenging task
to obtain an accurate estimate of the cross section.
A highly significant feature is that
the ratio of the 2B to
1B contribution for $hen$ is about $-4$,
which is much larger in magnitude
than the corresponding ratio $hep$, which is about $-0.6$.
This large difference
can be understood by recalling
the ``chiral filter mechanism" argument ~\cite{KDR}
according to which the $\nlo{1}$ contribution is non-vanishing
for $hen$, while the 2B correction in $hep$ only starts at
$\nlo{3}$.
It is remarkable that, despite this difference,
exactly the same MEEFT strategy works
for both $hep$ and $hen$.

The cutoff dependence of the  matrix elements
and the corresponding cross section are shown in
Table~\ref{tab:cutoff}. We remark that the 1B matrix element has no cutoff
dependence.
\begin{table}
\caption{\label{tab:cutoff} The cutoff dependence of the
matrix elements (in $10^{-3}$ fm$^{3/2}$)
and the corresponding total cross sections for
$\Lambda=$ 500, 600 and 800 MeV.}
\begin{ruledtabular}
\begin{tabular}{l|rrr}
$\Lambda$(MeV) &  500 & 600 & 800 \\
\hline
1B            & $-1.76(1)$& $-1.76(1)$& $-1.76(1)$ \\
\hline
2B(non-contact terms)& 5.24(2) & 6.79(3) & 8.31(3) \\
2B(contact terms) & 1.80(1) & 0.35(1) & $-0.99(1)$ \\
2B total       & 7.04(2) & 7.14(3) & 7.32(6) \\
\hline
1B+2B         & 5.29(2) & 5.39(3)  & 5.57(6) \\
\hline
$\sigma(\mu b)$& $56.9(3)$ & $59.2(5)$ & $63.2(10)$ \\
\end{tabular}
\end{ruledtabular}
\end{table}
It is noteworthy that the cutoff dependence seen in the $hen$ case
is quite similar to that seen in the $hep$ case. While the
individual contributions of the contact and non-contact terms
vary strongly as functions of $\Lambda$,
their sum shows a greatly reduced $\Lambda$-dependence.
This can be interpreted as a manifestation (albeit approximate)
of the renormalization group invariance of physical
observables.
The smallness of the cutoff dependence, $\sim 3\%$,
seen in Table~\ref{tab:cutoff}
indirectly indicates that our MEEFT scheme allows us to control
short-range dynamics to a satisfactory degree.
The remaining small cutoff dependence may be
attributed to the contributions of terms ignored in
this calculation, \eg, the ``fixed" terms, $n$-body currents for
$n>2$, other higher chiral order terms, etc.
Our final value for the threshold $hen$ cross section is
$\sigma = (60.1\pm 3.2\pm 1.0)~\mu b$,
where the first error comes from the cutoff dependence and the
second from the statistical errors.
Errors arising from the uncertainties in the
$n$-$^3$He scattering length are estimated to
be $\sim 1\%$ in the matrix element.

We have already mentioned that
the existing SNPA calculations for $hen$
cannot explain $\sigma_{exp}$.
A comment is in order on this point
since, for all the other cases so far studied
in both SNPA and MEEFT ($pp$ fusion, $hep$, $\nu-d$
scattering, radiative $np$-capture {\it etc}),
the numerical results exhibit a close resemblance
(except that the MEEFT results come with
systematic error estimates.)
As mentioned, one of the most important ingredients
of MEEFT is a ``renormalization" procedure in which the relevant unknown LECs
are fixed using the experimental values
of observables in neighboring nuclei.
A similar procedure has been done in the SNPA calculations
of the above-quoted cases.
%
However, the existing SNPA calculation of $hen$ \cite{hen1,hen2},
lacks this ``renormalization", and this explains
why the ``existing" SNPA calculation of $\sigma_{hen}$
disagrees with the experimental data.

To summarize, we have carried out an MEEFT calculation
for $hen$ up to $\nlo{3}$
and provided the first theoretical explanation of
the experimental value of the $hen$ cross section, $\sigma$.
Because of the drastic suppression of the
leading-order one-body term, which makes the sub-leading exchange-current
a dominant term, and
a significant cancellation between these two contributions,
an $unnatural$ amplification of the cutoff dependence occurs,
as in the $hep$ case.
Despite this accidental ``fine-tuning" situation,
we have shown that MEEFT allows us to make a parameter-free
estimation of $\sigma_{hen}$
with accuracy better than $\sim 10\%$.
The successful application of MEEFT to $hen$
renders strong support for the results of our previous MEEFT calculation for $hep$;
furthermore, it demonstrates the basic soundness of the MEEFT strategy in general.
The present treatment is open to several improvements
such as going to the next order in chiral perturbation,
using more accurate SNPA wave functions,
a more stringent control of mismatch in the
chiral counting between SNPA and a formally accurate chiral expansion that
enters in the currents, a better understanding of the role that
the counter terms play in the renormalization group property, etc.
It is reasonable, however, to expect
that the effects of these improvements
are essentially accommodated in the above-quoted error estimate based
on the cutoff dependence.
A robust estimation of the $hep$ S-factor
has been a long-standing challenging problem in nuclear physics~\cite{bahcall}.
We believe that our MEEFT calculation of $hep$ and $hen$
have solved this problem to a satisfactory degree.
Short of doing a fully formally consistent EFT calculation,
which is at present out of our reach, MEEFT seems the best one could do
at moment.

We would like to acknowledge invaluable discussions with, and
support from, Professors K. Kubodera, D.-P. Min and M. Rho with
whom this work was initiated.
The work of TSP was supported by
Korea Research Foundation Grant(KRF-2001-050-D00007).



\begin{thebibliography}{99}
\bibitem{hep} T.-S. Park \etal, Phys. Rev. {\bf C67}, 055206 (2003),
nucl-th/0208055.

\bibitem{BR03} G.E. Brown and M. Rho,
nucl-th/0305089, to appear in Phys. Rep.

\bibitem{Wolfs}
F.L.H. Wolfs \etal,
Phys. Rev. Lett. {\bf 63}, 2721 (1989).

\bibitem{wervelman}
R. Wervelman \etal, Nucl. Phys. {\bf A526}, 265 (1991).

\bibitem{hen1}
J. Carlson, D.O. Riska, R. Schiavilla and R.B. Wiringa, Phys. Rev. {\bf
C42}, 830 (1990).

\bibitem{hen2}
R. Schiavilla, R.B. Wiringa, V.R. Pandharipande, and J. Carlson,
Phys. Rev. {\bf C, 45} 2628 (1992).

\bibitem{zimmer} O. Zimmer \etal, EJPdirect {\bf A1}, 1 (2002) .

\bibitem{park2001} T.-S. Park, K. Kubodera, D.-P. Min, and M. Rho, Nucl. Phys. {\bf A684},
                   101 (2001).

\bibitem{weinberg} S. Weinberg, Phys. Lett. {\bf B251}, 288 (1990);
                   Nucl. Phys.{\bf B363},3 (1991);
                   Phys. Lett. {\bf B295}, 114 (1992).

\bibitem{excV} T.-S. Park, D.-P. Min and M. Rho, Phys. Rev. Lett. {\bf 74}, 4153 (1995);
Nucl. Phys. {\bf A596}, 515 (1996).

\bibitem{isoS} T.-S. Park, K. Kubodera, D.-P. Min, and M. Rho , Phys. Lett. {\bf B472}, 232 (2000)

\bibitem{kubodera} K. Kubodera, nucl-th/0308055.

\bibitem{stonybrook} S.K. Bogner, T.T.S. Kuo and A. Schwenk, Phys. Rep. {\bf 386}, 1 (2003).

\bibitem{fofman}H.M. Fofmann and G.M. Hale, Nucl. Phys. {\bf A613}, 69 (1997).


\bibitem{KDR} K. Kubodera, J. Delorme and M. Rho, Phys. Rev. Lett. {\bf 40}, 755 (1978);
M. Rho, Phys. Rev. Lett. {\bf 66}, 1275 (1991).

\bibitem{bahcall} J.N. Bahcall, Phys. Rep. {\bf 333}, 47 (2000),
 and references therein.

\end{thebibliography}
\end{document}